\documentclass{article}
\usepackage{ijcai17}

\usepackage{times}
\usepackage{epsfig}
\usepackage{graphicx}
\usepackage{amsmath}
\usepackage{amssymb}
\usepackage{multirow}
\usepackage{epstopdf}
\usepackage{graphicx}
\usepackage{subfigure}
\usepackage[small]{caption}

\title{Cross-modal Common Representation Learning by Hybrid Transfer Network}

\author{Xin Huang, Yuxin Peng\thanks{Corresponding author.},  \textmd{and} Mingkuan Yuan \\ 
Institute of Computer Science and Technology, 
Peking University, Beijing 100871, China \\
pengyuxin@pku.edu.cn}

\begin{document}

\maketitle

\begin{abstract}
DNN-based cross-modal retrieval is a research hotspot to retrieve across different modalities as image and text, but existing methods often face the challenge of insufficient cross-modal training data. In single-modal scenario, similar problem is usually relieved by transferring knowledge from large-scale auxiliary datasets (as ImageNet). 
Knowledge from such single-modal datasets is also very useful for cross-modal retrieval, which can provide rich general semantic information that can be shared across different modalities.
However, it is challenging to transfer useful knowledge from \textbf{\emph{single-modal}} (as image) source domain to \textbf{\emph{cross-modal}} (as image/text) target domain. 
Knowledge in source domain cannot be directly transferred to both two different modalities in target domain, and the inherent cross-modal correlation contained in target domain provides key hints for cross-modal retrieval which should be preserved during transfer process. 
This paper proposes Cross-modal Hybrid Transfer Network (CHTN) with two subnetworks: \emph{Modal-sharing transfer subnetwork} utilizes the modality in both source and target domains as a bridge, for transferring knowledge to both two modalities simultaneously; \emph{Layer-sharing correlation subnetwork} preserves the inherent cross-modal semantic correlation to further adapt to cross-modal retrieval task. Cross-modal data can be converted to common representation by CHTN for retrieval, and comprehensive experiments on 3 datasets show its effectiveness.

\end{abstract}

\section{Introduction}

With the rapid progress of human civilization and technology, multimodal data as image, text, video, and audio has been rapidly increasing on the Internet, and gradually becomes the main form of information. In this situation, cross-modal retrieval has become an important application of artificial intelligence, which is a novel paradigm where the retrieval results and user query are relevant in semantics but of different modalities. For example, user can submit an image to retrieve relevant text documents, and vice versa. 
Different from single-modal analysis as \cite{DBLP:conf/aaai/Peng15,DBLP:journals/tip/TangLWZ15}, cross-modal retrieval faces the challenge that 
 different modalities have inconsistent representations, and the existing mainstream is to learn common representation for them.
Recent years, cross-modal retrieval based on Deep Neural Network (DNN) has become an active research topic \cite{ngiam32011multimodal,srivastava2012learning,feng12014cross,DBLP:conf/ijcai/PengHQ16}, which aims to fulfill the DNN's strong ability of abstraction for dealing with complex cross-modal correlation.

Training data is important for the performance of DNN-based methods, but there is often insufficient training data for a specific task. In single-modal scenario, the problem of insufficient training data is usually relieved by the idea of transfer learning \cite{DBLP:journals/tkde/PanY10,DBLP:conf/ijcai/SamdaniY11,DBLP:conf/ijcai/ChenZ13}, which can transfer the knowledge of large-scale training data in source domain to target domain. For example, CNN model pre-trained on a subset of ImageNet \cite{DBLP:conf/cvpr/DengDSLL009} with over 1,200,000 labeled images usually acts as the basic model for many problems in computer vision \cite{DBLP:conf/nips/KrizhevskySH12}, which is from ImageNet large-scale visual recognition challenge (ILSVRC) 2012. However, the knowledge transfer is usually performed within the same modality. Knowledge contained in such large-scale single-modal datasets is also very valuable for cross-modal retrieval because it can provide rich general semantic information, which can be shared by different modalities to facilitate cross-modal semantic learning. But it is a challenging problem to effectively transfer useful knowledge from such \emph{single-modal source domain} to \emph{cross-modal target domain}. For example, we aim to train a cross-modal retrieval model on a small image/text dataset as target domain, and have a large-scale image dataset ImageNet as source domain. On the one hand, because there are only labeled images in ImageNet, the knowledge contained in source domain cannot be directly transferred to both image and text modalities in the target domain. On the other hand, the inherent cross-modal semantic correlation contained in the image/text dataset provides key hints for cross-modal retrieval and should be preserved in the transfer process.

For addressing the above problems, this paper proposes Cross-modal Hybrid Transfer Network (CHTN), which is a unified architecture consisting of two subnetworks: modal-sharing transfer subnetwork and layer-sharing correlation subnetwork. 
\emph{Modal-sharing transfer subnetwork} utilizes the modality contained in both source and target domains as a bridge, for propagating knowledge to both the two modalities in target domain simultaneously by single-modal transfer and cross-modal transfer.
\emph{Layer-sharing correlation subnetwork} focuses on preserving the inherent cross-modal semantic correlation to further adapt to the cross-modal retrieval task in the target domain.
CHTN performs knowledge transfer and correlation learning at the same time, and can generate common representation for different modalities to perform cross-modal retrieval.
For verifying the effectiveness of CHTN, we conduct cross-modal retrieval experiments on 3 widely-used datasets: Wikipedia, NUS-WIDE-10k, and Pascal Sentences. The experimental results show CHTN achieves the best accuracy among 10 state-of-the-art methods.

\section{Related Work}

\subsection{Cross-modal Retrieval}
Cross-modal retrieval is the basic research topic of this paper, which performs retrieval across different modalities as image and text. The current mainstream is common representation learning, which projects data of different modalities into a common space, and then the cross-modal similarity can be directly computed by distance measurement. 
The learning process is usually guided by co-existence (mainly pairwise correlation in the existing works) \cite{HotelingBiometrika36RelationBetweenTwoVariates,LiMM03CFA,ngiam32011multimodal,feng12014cross} and semantic correlation \cite{RasiwasiaMM10SemanticCCA,DBLP:journals/tmm/KangXLXP15,DBLP:conf/ijcai/PengHQ16}.  
According to the difference of basic models, existing works can be classified into traditional methods and DNN-based methods.

Traditional methods mainly learn linear projections for cross-modal common representation. Some representative methods are Canonical Correlation Analysis (CCA) \cite{HotelingBiometrika36RelationBetweenTwoVariates}, Cross-modal Factor Analysis (CFA) \cite{LiMM03CFA}. 
A few recent works take other information into consideration as semi-supervised regularization \cite{ZhaiTCSVT2014JRL} and local group based priori \cite{DBLP:journals/tmm/KangXLXP15}, but they are still based on linear projection. Recent years, inspired by the successful application of DNN in many single-modal tasks such as image classification, DNN-based methods for cross-modal retrieval have become an active research topic, which learn common representation with deep network as \cite{ngiam32011multimodal,kim2012learning,DBLP:conf/ijcai/WangCO015,DBLP:conf/ijcai/PengHQ16,DBLP:journals/tcyb/ZLWLZY17}. 

Existing DNN-based cross-modal methods mostly take hand-crafted feature vector as input, and are trained only with the target cross-modal datasets, so often face the challenge of insufficient training data. Some recent works as \cite{DBLP:journals/tcyb/WeiZLWLZY17} adopt CNN as a component in their network, where the CNN is pre-trained with a large-scale dataset (ImageNet here), and fine-tuned with the images in cross-modal dataset. That is to say, the knowledge from single-modal dataset (ImageNet) is only utilized to improve the image representation, but the knowledge transfer between different modalities is not involved. 
Our CHTN aims to make full advantage of the single-modal auxiliary dataset, by transferring semantic knowledge not only within the same modality, but also across different modalities simultaneously, so as to learn better cross-modal common representation and facilitate the performance of cross-modal retrieval.

\subsection{Transfer Learning}
Transfer learning \cite{DBLP:journals/tkde/PanY10} aims to propagate knowledge from source domain to target domain for relieving the problem of insufficient labeled training data, and has achieved success in wide applications \cite{DBLP:conf/ijcai/SamdaniY11,DBLP:conf/ijcai/ChenZ13}.
The idea of transfer learning plays a key role in DNN-based methods, because the performance of DNN-based methods often relies on training data size, but there is usually insufficient training data for specific tasks. For addressing this, there are some DNN-based transfer learning methods as \cite{DBLP:conf/icml/GlorotBB11,DBLP:conf/icml/LongC0J15}.

All the mentioned transfer learning technologies are limited in single-modal scenario, i.e., the source domain and target domain share the same single modality.
Beyond these, some works have been proposed involving more than one modality. For example, the method of \cite{DBLP:conf/cvpr/TsaiYW16} proposes heterogeneous transfer from one modality to another, which is still a one-to-one transfer paradigm. Some methods as \cite{DBLP:journals/tmm/YangZX15} perform knowledge transfer between two domains with two modalities, but they assume that the two domains both share the two modalities. 
The method proposed by \cite{DBLP:conf/cvpr/GuptaHM16} aims to transfer knowledge from a large-scale labeled image dataset to a dataset of paired images, where there are two different kinds of images (specifically, RGB image and Depth image). 

Our CHTN aims to address the knowledge transfer problem from \emph{single-modal source domain} (as image dataset) to \emph{cross-modal target domain} (as image/text dataset), which is different from the above mentioned works. Note that a recent work transitive hashing network (THN) \cite{DBLP:journals/corr/CaoL016a} has similar setting with our CHTN involving the knowledge transfer between single-modal and cross-modal datasets. However, THN and our CHTN aim at different problems with different focuses of method design: THN learns from an auxiliary cross-modal dataset to bridge two modalities from single-modal datasets; CHTN uses an auxiliary single-modal dataset to promote common representation learning in one cross-modal dataset.
This transfer paradigm can effectively utilize useful knowledge from single-modal dataset to relieve the problem of insufficient labeled cross-modal data for benefiting applications as cross-modal retrieval.

\begin{figure*}[t]
  \centering
\begin{minipage}[c]{\linewidth}
\centering
  \includegraphics[width=1\textwidth]{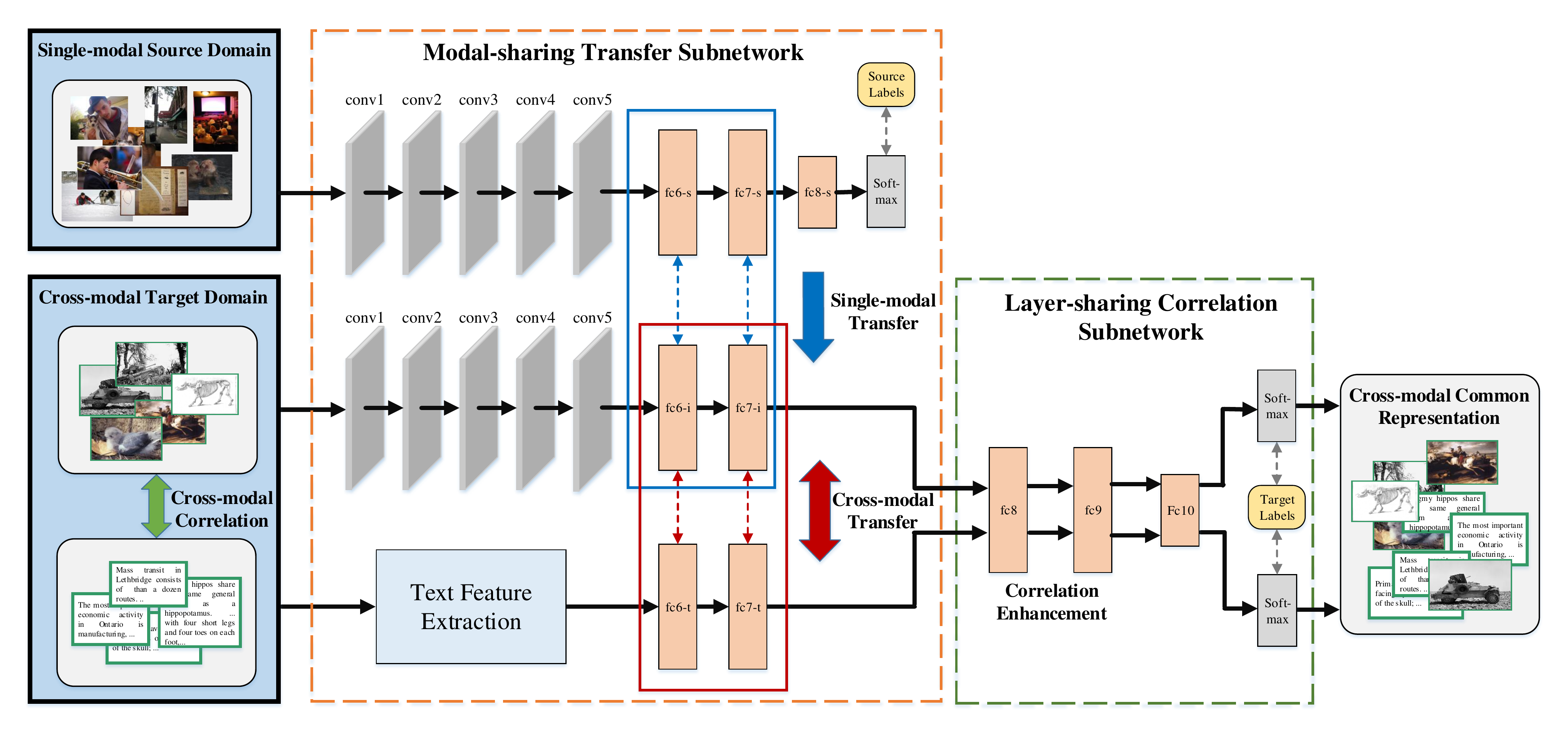}
\end{minipage}%
\caption{An overview of our Cross-modal Hybrid Transfer Network (CHTN).}\label{fig:network}
\end{figure*}

\section{Cross-modal Hybrid Transfer Network}

The overview of CHTN is as shown in Figure \ref{fig:network}, which is a unified architecture consisting of two subnetworks: \emph{modal-sharing transfer subnetwork} and \emph{layer-sharing correlation subnetwork}. In this paper, we take knowledge transfer from a large-scale image dataset (ImageNet) to an image/text dataset as an example to describe our methods.

We denote the single-modal source domain as $D^s=\left \{I^{s}\right \}$ with the labeled images $I^{s}=\left \{i_{r}^{s},y_{r}^{s}\right \}_{r=1}^{m}$,  $i_{r}^{s}$ is the $r$-th image in source domain and its label is $y_r^s$.
The cross-modal target domain is denoted as $D^t=\left \{I^{t},T^{t},I_u^{t},T_u^{t}\right \}$ with labeled images/text pairs $\{{I^{t},T^{t}}\}$ and unlabeled image/text pairs $\{{I_u^{t},T_u^{t}}\}$.  Similarly, $I^{t}=\left \{i_{p}^{t}, y_p^{t}\right \}_{p=1}^{n_l}$ denotes the labeled images in target domain, and $T^{t}=\left \{t_{q}^{t},y_{q}^{t}\right \}_{q=1}^{n_l}$ denotes the labeled text in the target domain. Correspondingly, the unlabeled images and text in target domain are denoted as $I_u^{t}=\left \{i_{p}^{t}\right \}_{p=n_l+1}^{n}$, and $T_u^{t}=\left \{t_{q}^{t}\right \}_{q=n_l+1}^{n}$. The aim is to transfer knowledge from $D^s$ to $D^t$, and learn cross-modal common representation as $R^{I}=\left \{r_{p}^{I}\right \}_{p=n_l+1}^{n}$ for unlabeled images and $R^{T}=\left \{r_{q}^{T}\right \}_{q=n_l+1}^{n}$ for unlabeled text in target domain. 


\subsection{Modal-sharing Transfer Subnetwork}
Modal-sharing transfer subnetwork is a hybrid architecture for transferring knowledge from image source domain to image/text target domain. It firstly uses convolutional layers to produce convolutional feature maps and receive representation of text (as BoW vector). Then the image feature maps and text representation will pass through fully-connected layers, where the knowledge transfer is performed. This subnetwork can be further viewed as two parts: single-modal knowledge transfer and cross-modal knowledge transfer, where the image pathway for target domain is shared by both the two parts, acting as a bridge for performing knowledge transfer from images in source domain to both the images and text in target domain.

{\bf Single-modal knowledge transfer.} 
It has been shown in many literatures that the model of DNN has considerable generalization ability. However, when we use DNN in a new domain, the domain discrepancy still exists \cite{yosinski2014transferable}.
To perform single-modal knowledge transfer from images in source domain to images in target domain, the key problem is the domain discrepancy between them. 

Following \cite{DBLP:conf/icml/LongC0J15}, we use feature adaptation method \cite{gretton2012kernel} which aims to minimize the Maximum Mean Discrepancy (MMD) of the same modality between source and target domains. By minimizing MMD, we can let a transferred model match the target domain distributions effectively, and so achieve the knowledge transfer within the same modality. We denote MMD between distribution $a$ of images from source domain $\{ i^s \}$ and distribution $b$ of images from target domain $\{ i^t \}$ by $d_k(a,b)$, so the squared formulation of MMD between $a$ and $b$ in reproducing kernel Hibert space (RKHS) $\mathcal{H}_k$ can be defined as:
\begin{align}
d_k^2(a,b)\overset{\Delta}{=}\left \| \mathbf{E}_a[\phi(i^s)]-\mathbf{E}_b[\phi(i^t)] \right \|^2_{\mathcal{H}_k}
\end{align}
where $\phi$ refers to the representation of a certain layer in deep neural network and the mean embedding of distribution $a$ in $\mathcal{H}_k$ is $\mu_k(a)$, so that $\mathbf{E}_{\textsc{x}\sim a}f(\textsc{x})=\left \langle f(\textsc{x}),\mu _k(a) \right \rangle_{\mathcal{H}_k}^2$ for all $f\in \mathcal{H}_k$. 
So the single-modal transfer loss can be written as:
\begin{align}
Loss_{Single}=  \sum_{l=l_6}^{l_7}d_k^2(I^s,I^t)
\end{align}
where the single-modal transfer loss at one layer is denoted as $d_k^2(I^s,I^t)$. $l_6$ and $l_7$ are corresponding layers in the network (i.e., fc6-s/fc6-i and fc7-s/fc7-i).

Moreover, the layers of source domain require fine-tuning itself on the source labeled instances. In this way, the labels of source domain can be used to provide supplementary supervision information and guide the single-modal knowledge transfer. The source domain supervision loss is as follows:
\begin{align}
Loss_{Source}=-\frac{1}{m}\sum_{r=1}^{m} f_s(i_r^s, y_r^s, \theta^s)
\end{align}
where $f_s(i_r^s, y_r^s, \theta^s)$ is the softmax loss function as:
\begin{align}
f_s(x, y, \theta) = \sum_{j=1}^c1\left \{ y=j \right \}\log \frac{e^{\theta_j x}}{\sum_{l=1}^c e^{\theta_l x}}
\end{align}
where $\theta$ are parameters of the network. $y$ denotes the label of instance $x$ and $c$ is the number of all possible classes of instance $x$. $1\left \{ y=j \right \}$ is an indicator function, which is equal to 1 if $y=j$, otherwise 0.

By minimizing the single-modal transfer loss and the source domain supervision loss, we can effectively decrease the domain discrepancy of image modality between the two domains, and perform single-modal knowledge transfer.

{\bf Cross-modal knowledge transfer.} 
The aforementioned single-modal knowledge transfer can only propagate knowledge from images in source domain to images in target domain. But because there exists cross-modal correlation between image and text in the target domain, it is possible to further achieve cross-modal knowledge transfer between image and text.

The image and text in an image/text pair are closely corresponded with each other in cross-modal dataset, so the knowledge can be effectively shared within each pair. 
The main idea of cross-modal knowledge transfer is to let the representations from high-level layers of pairwise data similar to each other, which is an intuitive idea as \cite{LiMM03CFA,feng12014cross}.
Specifically, we adopt Euclidean distance between cross-modal high-level layers (i.e., fc6-i/fc6-t and fc7-i/fc7-t) as the risk of cross-modal knowledge transfer, and denote the cross-modal discrepancy between an image instance $i_p^t$ and its paired text instance $t_p^t$ both from target domain as: 
\begin{align}
d_c^2(i_p^t,t_p^t)=\left \| \phi(i_p^t)-\phi(t_p^t) \right \|^2
\end{align} 
Then we get the cross-modal transfer loss as:
\begin{align}
Loss_{Cross}=\sum_{l=l_6}^{l_7}\sum_{p=1}^{n_l}d_c^2(i_p^t,t_p^t)
\end{align}
By minimizing this cross-modal transfer loss, we can achieve cross-modal knowledge transfer by reducing the discrepancy between representations of pairwise cross-modal data from high-level layers.

It should be noted that the modality shared by both source domain and target domain (image here) acts as a bridge to link the single-modal and cross-modal knowledge transfer. So these two kinds of transfer can be considered simultaneously, which leads to a hybrid knowledge transfer style. 
By optimization with Stochastic Gradient Descent (SGD), we can minimize these loss functions and perform knowledge transfer from single-modal source domain to cross-modal target domain in training stage. 

\subsection{Layer-sharing Correlation Subnetwork}

The inherent semantic correlation in cross-modal target domain is the key hint for bridging different modalities, which is the essential property of cross-modal dataset. Because our aim is to perform cross-modal retrieval, we further design a layer-sharing correlation subnetwork to enhance such correlation, and make the model more adapted to cross-modal retrieval task.

This subnetwork is a simple but effective task-specific structure for cross-modal retrieval, with layers shared by both image and text. 
As shown in Figure \ref{fig:network}, we use two shared fully-connected layers (fc8 and fc9) to construct it. Both the representations of image and text output from  modal-sharing transfer subnetwork will pass through the two fully-connected layers, and then there is a common classification layer.

Because the parameters of fc8 and fc9 are shared by both image and text, we can use the supervision information in cross-modal target domain to ensure the semantic correlation of different modalities. Given the labels of two paired modalities in target domain, the correlation loss is:
\begin{align}
Loss_{Correlation}= -\frac{1}{n_l}\sum_{p=1}^{n_l}( f_s(i_p^t, y_p^t, \theta^t) + f_s(t_p^t, y_p^t, \theta^t))
\end{align}
where $f_s$ is the softmax loss function in Equation 4. $f_s(i_p^t, y_p^t, \theta^t)$ is the target image supervision term, while  $f_s(t_p^t, y_p^t, \theta^t)$ is the corresponding target text supervision term.
After standard back-propagation with multiple iterations, we can minimize this correlation loss, and the semantic information in cross-modal target domain will be fully exploited for guiding cross-modal retrieval task. 

It should be noted that the modal-sharing subnetwork and layer-sharing correlation subnetwork are in a unified architecture, so the two subnetworks can be trained jointly and boost each other. Especially, the training of modal-sharing transfer network can also be guided by correlation loss, which makes the whole knowledge transfer processing further adapt to cross-modal retrieval task in target domain. In test stage, we take the predicted probability vectors as the final common representation $R^{I}$ and $R^{T}$ of image and text, for performing cross-modal retrieval.

\section{Experiments}

\subsection{Details of the Deep Architecture}

In this section, we present the details of our CHTN in the experiments. 
In the modal-sharing transfer subnetwork, there are three pathways: source image pathway, target image pathway, and target text pathway. In the source image pathway and target image pathway, we take five convolutional layers (conv1-conv5) of AlexNet \cite{DBLP:conf/nips/KrizhevskySH12} pre-trained on ImageNet from the Caffe\footnote{http://caffe.berkeleyvision.org} Model Zoo. It receives the images resized as $256 \times 256$ and generates convolutional feature maps (pool5) for images, and the five convolution layers are regarded as general layers to be frozen in training stage. All the three pathways of modal-sharing transfer subnetwork have two fully-connected layers, and all of them have $4,096 \times 4,096$ units. The base learning rates of all the fully-connected layers are set to be 0.01.

The MMD loss layers are implemented following \cite{DBLP:conf/icml/LongC0J15} between the two corresponding fully-connected layers of source image pathway and target image pathway, which performs single-modal knowledge transfer. Similarly, there are two contrastive loss layers from Caffe between the two fully-connected layers of target image pathway and target text pathway to realize cross-modal knowledge transfer. Moreover, because the magnitude of $Loss_{Cross}$ is much larger than those of $Loss_{Single}$, $Loss_{Source}$, and $Loss_{Correlation}$ (about 1,000 times), we set its weight as 0.001, and those of  $Loss_{Single}$, $Loss_{Source}$, and $Loss_{Correlation}$ are all 1. These parameter settings can be easily adjusted for other datasets in the implementation of network. After an fc8-s layer and a softmax loss layer of source image pathway, we can calculate softmax loss function for images of source domain in training stage. 

Layer-sharing correlation subnetwork consists of two fully-connected layers (fc8 and fc9), for which the base learning rates are set to be 0.01 and have $4,096 \times 4,096$ units. The two fully-connected layers receive the outputs from the last fully-connected layer of both target image pathway and target text pathway in modal-sharing transfer subnetwork. Finally, after a fully-connected layer and a softmax layer, we can calculate softmax loss function for images and text of target domain in training stage, and get the probability vector as common representation in test stage.

\subsection{Dataset Introduction}

In the experiments, ImageNet serves as the single-modal source domain, and we adopt a widely-used subset with over 1,200,000 labeled images \cite{DBLP:conf/nips/KrizhevskySH12} from ILSVRC 2012.
We perform knowledge transfer from ImageNet to 3 cross-modal datasets as target domains respectively, and conduct cross-modal retrieval on them, namely Wikipedia, NUS-WIDE-10k and Pascal Sentences. For fair comparison, we strictly take the same  dataset partition according to \cite{feng12014cross,DBLP:conf/ijcai/PengHQ16} for our CHTN and all the compared methods in the experiments. 

{\textbf {Wikipedia dataset}} \cite{RasiwasiaMM10SemanticCCA} is the most widely-used dataset for cross-modal retrieval evaluation as \cite{ZhaiTCSVT2014JRL,feng12014cross,DBLP:conf/ijcai/PengHQ16}, which is constructed from the Wikipedia ``featured articles''. This dataset has 2,866 image/text pairs classified into 10 high-level semantic categories: art, biology, geography, history, literature, media, music, royalty, sport, and warfare. In each image/text pair, the text contains several paragraphs of description for the image, so they have close correlation, and each pair exclusively belongs to one of the 10 categories. The dataset is randomly split into training set with 2,173 pairs, test set with 462 pairs, and validation set with 231 pairs. 

{\textbf {NUS-WIDE-10k dataset}} \cite{feng12014cross} is a subset of NUS-WIDE dataset \cite{NUSWIDE}. NUS-WIDE dataset is a relatively large-scale image/tag dataset with about 270,000 images. Each image has several corresponding text tags, which are used as the text modality in the experiments. NUS-WIDE dataset has 81 categories, but there exist overlaps among the categories. NUS-WIDE-10k dataset is constructed by selecting 10,000 image/text pairs evenly from 10 largest categories of NUS-WIDE dataset, under conditions that each pair exclusively belongs to one of the 10 categories, so there are 1,000 pairs for each category.
The dataset is randomly split into training set with 8,000 pairs, test set with 1,000 pairs, and validation set with 1,000 pairs evenly from the 10 categories.

{\textbf {Pascal Sentences dataset}} \cite{PascalSentence} is also an image/text dataset which is selected from 2008 PASCAL development kit. This dataset has 1,000 images evenly belonging to 20 categories, and each image has 5 description sentences. Similar to NUS-WIDE-10k dataset, Pascal Sentences dataset is randomly split into training set with 800 pairs, test set with 100 pairs, and validation set with 100 pairs evenly from the 20 categories.

It should be noted that among the compared methods, CMDN, Corr-AE, Bimodal AE, and Multimodal DBN need validation set for parameter adjustment, while the other methods including our CHTN don't. So for the methods except CMDN, Corr-AE, Bimodal AE, and Multimodal DBN, validation set is not used for the whole training and test stages.

\subsection{Evaluation Metrics}

In the experiments, two retrieval tasks are conducted: retrieving text by image query (Image$\rightarrow$Text) and retrieving images by text query (Text$\rightarrow$Image). Specifically, we first obtain the common representation for the images and text in the test set with CHTN and all the compared methods. Then we take one of the images in test set as query, compute the cross-modal similarity with all text in test set by cosine distance, and evaluate the ranking list by mean average precision (MAP), and vice-versa. It should be noted that the MAP score is computed for \emph{all the retrieval results} for comprehensive evaluation.

The MAP scores are computed as the mean of average precision (AP) for all queries, and AP is computed as:
\begin{align}
   AP = \frac 1 R \sum_{k=1}^n \frac {R_k} k \times rel_k
\end{align}
where $R$ denotes relevant item number in test set (according to the label in our experiments), $R_k$ denotes the relevant item number in top $k$ results, $n$ denotes the test set size, and $rel_k = 1$ means the $k$-th result is relevant, and 0 otherwise.


\subsection{Compared Methods and Input Settings}

Totally 10 state-of-the-art methods are compared in the experiments: CCA \cite{HotelingBiometrika36RelationBetweenTwoVariates}, CFA \cite{LiMM03CFA}, KCCA (with Gaussian kernel and polynomial kernel) \cite{DBLP:journals/neco/HardoonSS04}, Bimodal AE \cite{ngiam32011multimodal}, Multimodal DBN \cite{srivastava2012learning}, Corr-AE \cite{feng12014cross}, JRL \cite{ZhaiTCSVT2014JRL}, LGCFL \cite{DBLP:journals/tmm/KangXLXP15}, CMDN \cite{DBLP:conf/ijcai/PengHQ16} and Deep-SM \cite{DBLP:journals/tcyb/WeiZLWLZY17}. Among these, CCA, CFA, KCCA, JRL, LGCFL are traditional methods, and Bimodal-AE, Multimodal DBN, Corr-AE, CMDN and Deep-SM are DNN-based methods.

For image, CHTN has the end-to-end processing ability and directly takes the image pixels as input. However, all the compared methods except Deep-SM and our CHTN can only take extracted feature vector as input. So for fair comparison, we use the same AlexNet adopted by CHTN, and further fine-tuned to convergence with the images in each dataset as feature extractor for them. 
For text, we take exactly the same input feature as \cite{feng12014cross,DBLP:conf/ijcai/PengHQ16} for all compared methods. In detail, the input feature is 3,000 dimensional BoW vector for Wikipedia dataset, and 1,000 dimensional BoW vector for both NUS-WIDE-10k and Pascal Sentences datasets.

\subsection{Experimental Results}
\begin{figure}[t]
  \centering
\begin{minipage}[c]{\linewidth}
\centering
  \includegraphics[width=1\textwidth]{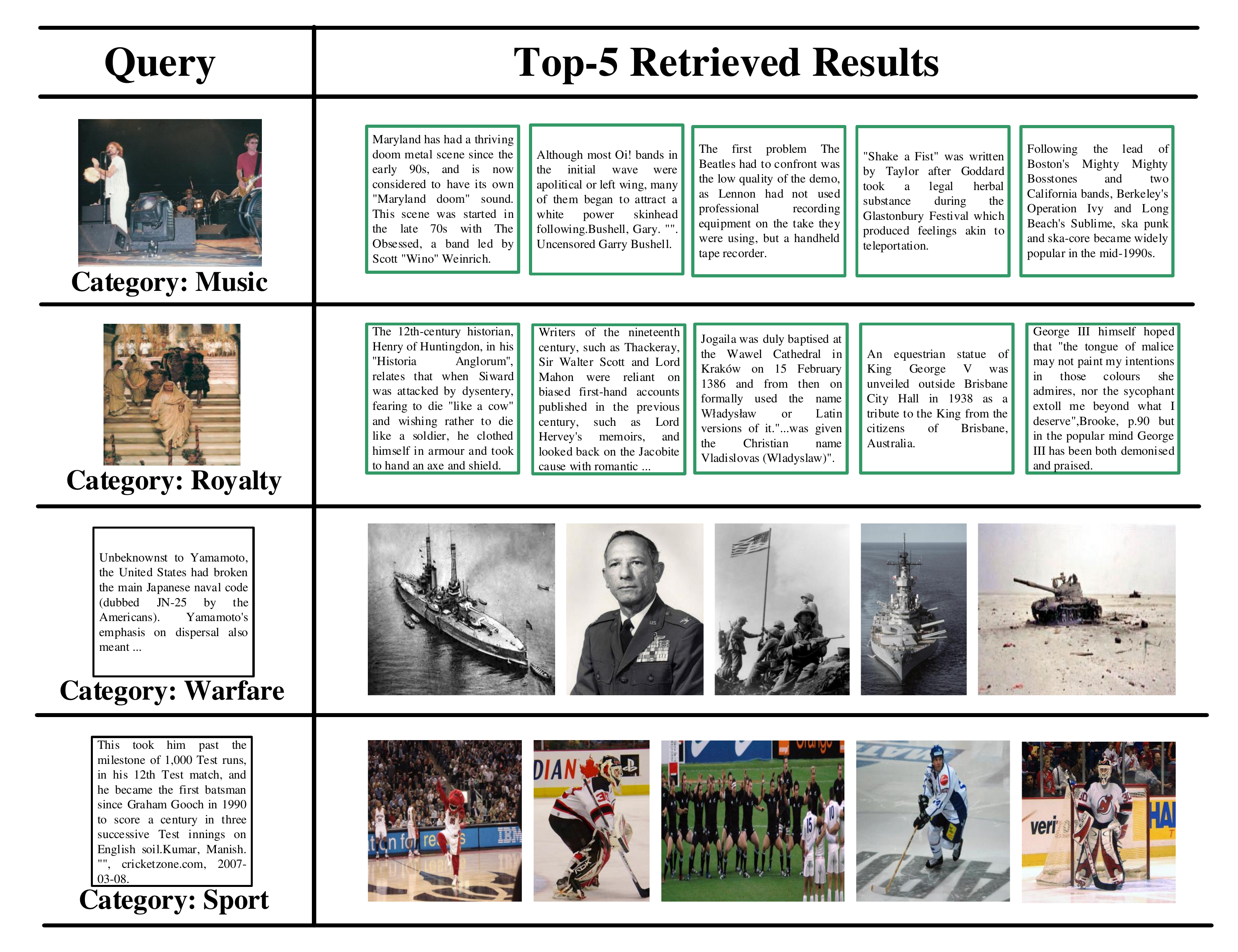}
\end{minipage}%
\caption{Some examples of cross-modal retrieval results on Wikipedia dataset obtained by CHTN. All the top-5 retrieval results are correct.}\label{fig:WikiEx}

\end{figure}

Table \ref{table:Results} shows the MAP scores of our CHTN and the compared methods on the 3 datasets. It can be seen that on Wikipedia dataset, CHTN achieves an inspiring accuracy improvement from 0.402 to 0.470, compared with the state-of-the-art method Deep-SM. Figure \ref{fig:WikiEx} shows some examples of cross-modal retrieval results on Wikipedia dataset. On NUS-WIDE-10k dataset, CHTN keeps the best and achieves the MAP score of 0.517. The result trend on Pascal Sentences dataset is much different from Wikipedia and NUS-WIDE-10k datasets. On Pascal Sentences dataset, the compared methods Bimodal AE and Multimodal DBN achieve similar accuracy with Deep-SM, and CMDN obtains the best accuracy among all compared methods. However, the accuracy of CHTN is still the highest. 
It can be also seen that the performance of compared methods is not stable.
For example, on Wikipedia dataset CMDN has a very clear advantage over Multimodal DBN, but on Pascal Sentences dataset they are comparable. Our CHTN keeps the best on the 3 datasets, which shows the hybrid transfer paradigm has generality, and it effectively transfers the knowledge from single-modal source domain to cross-modal target domain for learning better common representation and improving the accuracy of cross-modal retrieval.

 \begin{table}[htb]
 \begin{center}
 \tiny
 \begin{tabular}{|c|c|c|c|c|} 
 \hline
 \multirow{2}{*}{Dataset}&
 \multirow{2}{*}{Method} & \multicolumn{3}{c|}{Task}\\
 \cline{3-5}
  & & Image$\rightarrow$Text & Text$\rightarrow$Image & Average \\
  \hline
  
  \multirow{12}{*}{\begin{tabular}{c} Wikipedia \\ dataset \end{tabular}} &  CCA & 0.176 & 0.178 & 0.177 \\
   &  CFA & 0.330 & 0.306& 0.318 \\
   &  KCCA(Poly) & 0.230 & 0.224& 0.227\\
   &  KCCA(Gaussian) & 0.357 & 0.328& 0.343 \\
   &  Bimodal AE & 0.301 & 0.267& 0.284 \\
   &  Multimodal DBN & 0.204 & 0.145& 0.175 \\
   &  Corr-AE & 0.373 & 0.357& 0.365 \\
   &  JRL & 0.408 & 0.353& 0.381 \\
   &  LGCFL & 0.416 & 0.360& 0.388 \\
   &  CMDN & 0.409 & 0.364& 0.387 \\
   &  Deep-SM & 0.458 & 0.345& 0.402 \\
   &  \textbf{our CHTN} & \textbf{0.508} & \textbf{0.432} & \textbf{0.470} \\
  \hline
 
  \multirow{12}{*}{\begin{tabular}{c} NUS-WIDE\\ -10k \\dataset \end{tabular}} &  CCA & 0.159 & 0.189& 0.174 \\
   &  CFA & 0.299 & 0.301& 0.300 \\
   &  KCCA(Poly) & 0.129 & 0.157& 0.143\\
   &  KCCA(Gaussian) & 0.295 & 0.162& 0.229 \\
   &  Bimodal AE & 0.234 & 0.376& 0.305 \\
   &  Multimodal DBN & 0.178 & 0.144& 0.161 \\
   &  Corr-AE & 0.306 & 0.340& 0.323 \\
   &  JRL & 0.410 & 0.444& 0.427 \\
   &  LGCFL & 0.408 & 0.374& 0.391 \\
   &  CMDN & 0.410 & 0.450& 0.430 \\
   &  Deep-SM & 0.389 & 0.496& 0.443 \\
   &  \textbf{our CHTN} & \textbf{0.518} & \textbf{0.516} & \textbf{0.517} \\
  \hline
  
   \multirow{12}{*}{\begin{tabular}{c} Pascal \\ Sentences \\ dataset \end{tabular}} &  CCA & 0.110 & 0.116& 0.113 \\
    &  CFA & 0.341 & 0.308& 0.325 \\
    &  KCCA(Poly) & 0.271 & 0.280& 0.276\\
    &  KCCA(Gaussian) & 0.312 & 0.329& 0.321 \\
    &  Bimodal AE & 0.404 & 0.447& 0.426 \\
    &  Multimodal DBN & 0.438 & 0.363& 0.401 \\
    &  Corr-AE & 0.411 & 0.475& 0.443 \\
    &  JRL & 0.416 & 0.377 & 0.397 \\
    &  LGCFL & 0.381 & 0.435& 0.408 \\
    &  CMDN & 0.458 & 0.444& 0.451 \\
    &  Deep-SM & 0.440 & 0.414& 0.427 \\
    &  \textbf{our CHTN} & \textbf{0.467} & \textbf{0.477} & \textbf{0.472} \\
   \hline
 \end{tabular} 
 
 \end{center}
 \caption{MAP scores of our CHTN and compared methods.}
 \label{table:Results}
 \end{table}

 \begin{table}[htb]
 \begin{center}
 \tiny
 \begin{tabular}{|c|c|c|c|c|} 
 \hline
 \multirow{2}{*}{Dataset}&
 \multirow{2}{*}{Method} & \multicolumn{3}{c|}{Task}\\
 \cline{3-5}
  & & Image$\rightarrow$Text & Text$\rightarrow$Image & Average \\
  \hline
  
  \multirow{4}{*}{\begin{tabular}{c} Wikipedia \\ dataset \end{tabular}} &   \multirow{1}{*}{\begin{tabular}{c} CHTN (OnlyCross) \end{tabular}} & \multirow{1}{*}{0.465} & \multirow{1}{*}{0.407}& \multirow{1}{*}{0.436} \\
  & \multirow{1}{*}{\begin{tabular}{c} CHTN (NoShare) \end{tabular}} & \multirow{1}{*}{0.483} & \multirow{1}{*}{0.422}& \multirow{1}{*}{0.453} \\
    & \multirow{1}{*}{\begin{tabular}{c} CHTN (NoSrcSp) \end{tabular}} & \multirow{1}{*}{0.489} & \multirow{1}{*}{0.415}& \multirow{1}{*}{0.452} \\
   & \textbf{our CHTN} & \textbf{0.508} & \textbf{0.432} & \textbf{0.470}  \\
  \hline
  
  \multirow{4}{*}{\begin{tabular}{c} NUS-WIDE \\-10k \\ dataset \end{tabular}} &   \multirow{1}{*}{\begin{tabular}{c} CHTN (OnlyCross) \end{tabular}} & \multirow{1}{*}{0.360} & \multirow{1}{*}{0.406}& \multirow{1}{*}{0.383} \\
  & \multirow{1}{*}{\begin{tabular}{c} CHTN (NoShare) \end{tabular}} & \multirow{1}{*}{0.488} & \multirow{1}{*}{0.442}& \multirow{1}{*}{0.465} \\
   & \multirow{1}{*}{\begin{tabular}{c} CHTN (NoSrcSp) \end{tabular}} & \multirow{1}{*}{0.487} & \multirow{1}{*}{0.495}& \multirow{1}{*}{0.491} \\
   & \textbf{our CHTN} & \textbf{0.518} & \textbf{0.516} & \textbf{0.517}  \\
  \hline
   
  \multirow{4}{*}{\begin{tabular}{c} Pascal \\Sentences \\ dataset \end{tabular}} &   \multirow{1}{*}{\begin{tabular}{c} CHTN (OnlyCross) \end{tabular}} & \multirow{1}{*}{0.433} & \multirow{1}{*}{0.443}& \multirow{1}{*}{0.438} \\
  & \multirow{1}{*}{\begin{tabular}{c} CHTN (NoShare) \end{tabular}} & \multirow{1}{*}{0.446} & \multirow{1}{*}{0.467}& \multirow{1}{*}{0.457} \\
   & \multirow{1}{*}{\begin{tabular}{c} CHTN (NoSrcSp) \end{tabular}} & \multirow{1}{*}{0.435} & \multirow{1}{*}{0.457}& \multirow{1}{*}{0.446} \\
   &  \textbf{our CHTN} & \textbf{0.467} & \textbf{0.477} & \textbf{0.472} \\
  \hline
  
 \end{tabular} 
 
 \end{center}
 \caption{MAP scores of our CHTN and the baselines.}
 \label{table:Baseline}
 \end{table}

Table \ref{table:Baseline} shows the MAP scores of our baselines and the complete CHTN. CHTN (OnlyCross) means we remove the source image pathway in modal-sharing transfer subnetwork, and there is no layer-sharing correlation subnetwork (the fc7-i and fc7-t layers are directly connected to classification layers). This is a very basic and intuitive way to perform cross-modal retrieval learning with pairwise correlation and semantic information.
CHTN (NoShare) means there is no layer-sharing correlation subnetwork, and the fc7-i and fc7-t are directly connected to classification layers.
CHTN (NoSrcSp) means there is no source supervision loss of Eqn. (3), which is designed to verify if the supervision information of source domain can provide complementary semantic information for the target domain.
Except for the above differences, the rest parts of the three baselines keep the same with complete CHTN.

By comparing CHTN (OnlyCross) and CHTN (NoShare), we can see the single-modal transfer part provides important supplementary information to cross-modal transfer part, which leads to better accuracy.
By comparing CHTN (NoShare) with complete CHTN, we can see the preserving of inherent cross-modal semantic correlation helps the deep model further adapt to cross-modal retrieval task in target domain. By comparing CHTN (NoSrcSp) with complete CHTN, we can see that although label space and task of the two domains may be different, they can share relevant high-level semantic knowledge for improving retrieval accuracy.
The above baseline experiments show the modal-sharing transfer architecture and layer-sharing correlation subnetwork both play important role in our CHTN, and they can boost each other for better accuracy.


\section{Conclusion}

In this paper, we have proposed a Cross-modal Hybrid Transfer Network (CHTN), for addressing the problem of knowledge transfer from single-modal source domain (as image) to cross-modal target domain (as image/text). CHTN can be viewed as a unified model with two subnetworks: Modal-sharing transfer subnetwork utilizes the modality in both source and target domains as a bridge, for transferring knowledge to both two modalities simultaneously; Layer-sharing correlation subnetwork focuses on preserving the inherent cross-modal semantic correlation to further adapt to cross-modal retrieval task in target domain.
Experiments show that CHTN effectively utilizes the large-scale single-modal dataset to improve the cross-modal retrieval accuracy. 
The future works lie in two aspects. First, we intend to apply CHTN to other cross-modal applications like image caption to verify its generality. Second, we will focus on how to transfer knowledge from single-modal source domain to target domain with more than two modalities. 

\section*{Acknowledgments}
This work was supported by National Natural Science Foundation of China under Grants 61371128 and 61532005.

\appendix

\bibliographystyle{named}
\bibliography{ijcai17}

\end{document}